\documentclass[traditabstract,onecolumn]{aa}
\usepackage{txfonts}
\usepackage{pslatex}
\usepackage{caption}
\usepackage{graphicx}
\usepackage{dcolumn}
\usepackage{bm}
\usepackage{subfig}

\newcommand\simlt{\lower.5ex\hbox{$\; \buildrel < \over \sim \;$}}
\usepackage[numbers]{natbib}
\bibpunct{}{}{;}{n}{}{,} 
\begin{document}

\title{Deep observation of the giant radio lobes of Centaurus A with the Fermi Large Area Telescope}
\author{Rui-zhi Yang\inst{1, 2}
\and Narek Sahakyan\inst{3,4}
\and Emma de Ona Wilhelmi\inst{1}
\and Felix Aharonian\inst{1,5}
\and Frank Rieger\inst{1}
}
\institute{
 Max-Planck-Institut f{\"u}r Kernphysik, P.O. Box 103980, 69029 Heidelberg, Germany
\and Key Laboratory of Dark Matter and Space Astronomy, Purple Mountain Observatory, Chinese Academy of Sciences, Nanjing, 210008, China
\and University of Rome Sapienza and ICRANet, Dip. Fisica, p.le A. Moro 2, 00185 - Rome, Italy
\and Institute for Physical Research, NAS of Armenia, Ashtarak-2, 0203, Armenia
\and Dublin Institute for Advanced Studies, 31 Fitzwilliam Place, Dublin 2, Ireland
}%

\begin{abstract}
{The detection of high-energy (HE) $\gamma$-ray emission up to $\sim 3$ GeV from the giant lobes of the radio galaxy
Centaurus A has been recently reported by the Fermi-LAT Collaboration based on ten months of all-sky survey observations.
A data set more than three times larger is used here to study the morphology and photon spectrum of the lobes with higher statistics.
The larger data set results in the detection of HE $\gamma$-ray emission (up to $\sim 6$ GeV) from the lobes with a significance of
more than $10$ and $20~\sigma$ for the north and the south lobe, respectively. Based on a detailed spatial analysis and comparison
with the associated radio lobes, we report evidence for a substantial extension of the HE $\gamma$-ray emission beyond the WMAP
radio image for the northern lobe of Cen A. We reconstructed the spectral energy distribution (SED) of the lobes using radio
(WMAP) and Fermi-LAT data from the same integration region. The implications are discussed in the context of hadronic and time-dependent leptonic scenarios.}
\end{abstract}
\maketitle

\section{Introduction}
The bright, nearby radio galaxy Centaurus A (Cen~A; NGC 5128) has been extensively  studied from radio to very-high-energy (VHE)
$\gamma$-rays (e.g., see \cite{israel98, steinle10} for reviews). Its unique proximity (d$\sim$3.7~Mpc; \cite{ferrarese07}) and peculiar
morphology allow a detailed investigation of the non-thermal acceleration and radiation processes occurring in its active
nucleus and its relativistic outflows. At radio frequencies Cen~A reveals giant structures, the so-called "lobes", with a total angular
size of $\sim 10^{\circ}$ (\cite{shain58, burns83}), corresponding to a physical extension of $\sim 600~{\rm kpc}~ (\rm d/3.7 \rm~ Mpc)$.\\
At high-energy (HE; 200 MeV$<\rm E<100$ GeV) Fermi-LAT has recently detected $\gamma$-ray emission from the core (i.e.,
within \mbox{$\sim 0.1^{\circ}$}) and the giant radio lobes of Cen A (\cite{abdo10a, abdo10b}): An analysis of the available
ten-month data set reveals a point-like emission region coincident with the position of the radio core of Cen A, and two large extended
emission regions detected with a significance of $5$ and $8\sigma$ for the northern and the southern lobe, respectively. The HE
emission from the core extends up to $\sim 10$ GeV and is well described by a power-law function with photon index \mbox{$\sim2.7$}.
It can be successfully interpreted as originating from synchrotron self-Compton (SSC) processes in the innermost part of the relativistic
jet. However, a simple extrapolation of the HE core spectrum to the TeV regime tends to under-predict the TeV flux observed by
H.E.S.S. (\cite{aharonian09}), which may indicate an additional contribution related to, e.g. non-thermal magnetospheric
processes emerging at the highest energies (see \cite{rieger11} for review). The extended HE emission regions, on the other hand, seem to be
morphologically correlated with the giant radio lobes and contribute more than one-half to the total HE source emission. These
regions are again spectrally well described by a power-law function extending up to 2 or 3 GeV with photon indices of $\Gamma\sim
2.6$.\\

If the extended HE emission is generated by indeed  inverse-Compton up-scattering of CMB and EBL (extragalactic background light)
photons, this could offer a unique possibility to spatially map the underlying relativistic electron distribution in this source.
The detection of GeV $\gamma$-rays from the radio lobes implies magnetic field strengths $\simlt 1~\mu$G (e.g., \cite{abdo10a}).
This estimate can be obtained quite straightforwardly from the comparison of radio and $\gamma$-rays, assuming that these radiation
components are produced in the same region by the same population of electrons through synchrotron and inverse-Compton processes.
In general, however, the radio and the $\gamma$-ray region do not need to coincide. While the radio luminosity depends on the product
of the relativistic electron density $\rm N_e$ and the magnetic-field square $\rm B^2$, the inverse-Compton $\gamma$-ray luminosity
only depends on $\rm N_e$. This implies that $\gamma$-rays can give us model-independent information about both the energy
and the spatial distribution of electrons, while the radio image of synchrotron radiation strongly depends on the magnetic field. As
a consequence, the $\gamma$-ray image can be larger than the radio image if the magnetic field drops at the periphery of the region
occupied by electrons. This provides one of the motivations for a deeper study of the extended HE (lobe) emission region in Cen~A.
In principle, X-ray observations could also offer valueable insights into the electron (albeit in a different energy band) and the
magnetic field distributions in the lobes (e.g., \cite{hardcastle05}). Gamma-ray observations, however, bear a certain advantage in that
the characteristic cooling timescales for electrons that upscatter CMB photons into the GeV domain is much shorter ($t_c \sim 10^6 [10^6/
\gamma]$ yr), thereby enabling one to trace local conditions much more closely. Since the TeV electrons responsible for GeV $\gamma$-rays
cannot move much beyond ($\simlt$1 kpc) their acceleration points, the very fact of the existence of extended $\gamma$-ray lobes in
Cen~A implies that we deal with a huge (distributed) 200 kpc-size TeV electron accelerator. Moreover, for a hard uncooled spectrum
of low-energy electrons, the inverse Compton emission in the X-ray band will be suppressed. Therefore, an X-ray detection of the very
extended lobes is complicated (cf. \cite{isobe01}).

 In the present paper we analyze 3 yr of Fermi LAT data, increasing the available observation time by more than a factor of three with
respect to the previously reported results. The larger data set allows a detailed investigation of the spectrum and morphology of the
lobes with better statistics, especially above $1$ GeV, where the spectral shape may reflect cooling effects and/or maximum energy
constraints on the parent population of particles generating the HE $\gamma$-ray emission. We also re-analyze radio data from WMAP (\cite{page03}) for the same region from where the HE emission is evaluated from, and discuss the implications of the resulting spectral
energy distribution (SED) for different emission scenarios.\\
The paper is structured as follows. In Sec. 2 the spectral and spatial HE analysis results are described, while the analysis of the
WMAP results for the lobes is presented in Sec. 3. Implications for leptonic and hadronic emission models are discussed in Sec. 4
and conclusions are presented in Sec. 5.

\section{Fermi-LAT data analysis}
The Large Area Telescope (LAT) on board the Fermi $\gamma$-Ray Space Telescope, operating since August 4, 2008, can detect
$\gamma-$ray photons with energies in the range between $100$ MeV and a few $100$ GeV. Details about the LAT instrument can
be found in Atwood et al. (2009).  Here we analyze the field of view (FoV) of Cen~A, which includes the bright core and the giant
radio lobes. We selected data obtained from the beginning of the operation until November 14, 2011, amounting to $\sim 3$ yr of
data (MET 239557417-- 342956687). We used the standard LAT analysis software (v9r23p1)\footnote{http://fermi.gsfc.nasa.gov/ssc}.
To avoid systematic errors due to poor determination of the effective area at low energies, we selected only events with
energies above 200 MeV. The region-of-interest (ROI) was selected to be a rectangular region of size $14^ \circ \times 14^ \circ$
centered on the position of Cen A ($\rm RA=201^{\circ}21^{\prime}54^{\prime\prime}, \rm DEC=-43^{\circ}1^{\prime}9^{\prime\prime}$).
To reduce the effect of Earth albedo backgrounds, time intervals when the Earth was appreciably in the FoV (specifically, when
the center of the FoV was more than $52^ \circ$ from zenith) as well as time intervals when parts of the ROI were observed at zenith
angles $> 105^\circ$ were also excluded from the analysis. The spectral analysis was performed based on the P7v6 version of the
post-launch instrument response functions (IRFs). We modeled the Galactic background component using the LAT standard diffuse
background model {\it gal\_2yearp7v6\_v0} and we left the overall normalization and index as free parameters. We also used
{\it iso\_p7v6source} as the isotropic $\gamma$-ray background.\\
The resulting Fermi-LAT counts map for the 3 yr  data set is shown in Fig. 1(a). The (green) crosses show the position of the point-like
sources from the 2FGL catalog (\cite{abdo11}) within the ROI. The core of Cen A is clearly visible with a test statistic of TS $>$
800, corresponding to a detection significance of 28 $\sigma$. Extended emission to the north and south of Cen A is detected with
significances of TS $>100$ (10 $\sigma$) and TS $>400$ (20 $\sigma$), respectively.

\subsection{Spatial analysis}
Events with energies between $200$ MeV and $30$ GeV were selected. The residual image after subtracting the diffuse background
and point-like sources including the core of Cen A is shown in Fig. 1(b). The fluxes and spectral indices of 11 other point-like sources
generated from the 2FGL catalog within the ROI are also left as free parameters in the analysis. The 2FGL catalog source positions are
shown in Fig.~1(a), where 2FGL J1324.0-4330e accounts for the lobes (both north and south).
A new point-like source (2FGL J1335.3-4058), located at $\rm RA=203^{\circ}49^{\prime}30^{\prime\prime}, \rm DEC=-40^{\circ}34^{\prime}
48^{\prime\prime}$ accounts for some residual emission from the north lobe, although no known source at other wavelengths is found
to be associated. We treat it as part of the north lobe here. The core of Cen~A is modeled as a point-like source. Then the following
steps were performed:\\
(1) To evaluate the total (extended) HE $\gamma$-ray emission we first used a template based on the residual map (T1; corresponding
to the blue contours in Fig. 2). The TS values for the south and the north lobe in this template are 411 and 155, respectively. The residual
map was also compared with radio (WMAP, 22~GHz) lobe contours (green contours overlaid on Fig. 1(b)).  While lower-frequency radio
maps exist, we expect the higher-frequency 22 GHz map to better represent the GeV-emitting particles. We find that the south lobe of
the HE $\gamma$-ray image is similar to the south lobe of the radio one, whereas the HE emission in the north extends beyond the
radio lobe emission region. \\
(2) To understand this feature better, we re-fitted the excess using an additional template (T2; red contours in Fig.~3) generated
from the radio (WMAP) image. The two templates are shown in Fig.~2, and the corresponding residual maps are shown in Fig.~1.
While there is some residual emission to the north of Cen~A for template T2, this residual emission is obviously absent from template T1.
The qualitative features of the different residual maps are confirmed by the corresponding TS values, which are listed in Table 1.
Accordingly, the HE south lobe seems to agree reasonably well with the radio south lobe, whereas for the north lobe, the template
generated from the radio lobe (T2) fits the HE excess substantially worse than T1 (110 vs 155).

\begin{table}[htbp]
\caption{TS value for the two templates used.} \label{tab:1} \centering
\begin{tabular}{llll}
\hline
Model &\vline north Lobe&\vline south Lobe\\
\hline
T1   &\vline155 &\vline411\\

\hline
T2   &\vline110 &\vline406\\
\hline
\end{tabular}
\end{table}

(3) To further investigate a possible extension (or contribution of a background source) of the north lobe, we evaluated the projection
of a rectangular region on the excess image (in white in Fig. 1(b)). Fig. 3 shows the projection for the north and south regions for 
Fermi-LAT image (in black) and the radio one (in red). The south projection for the radio map is well-fitted by a single Gaussian centered
at $\sim 0.05$ (0 is defined as the center of the rectangle on $RA=201^{\circ}21^{\prime}54^{\prime\prime}, DEC=-43^{\circ}1^{\prime}
9^{\prime\prime}$) with an extension of $\sigma = 0.99^{\circ}$. For the Fermi-LAT map the Gaussian is centered at $\sim 0.5$ and has
$\sigma = 1.01^{\circ}$, compatible with the radio map projection. In contrast, the north projection for the Fermi-LAT map has a
Gaussian profile with $\sigma = 1.68^{\circ}$, while for the radio map $\sigma$ is $0.97^{\circ}$. The extension in the north projections
for the Fermi-LAT map indicates that the $\gamma$-ray north lobe is more extended than the radio one or that an (otherwise unknown)
source in the background may be contributing to the total emission. \\

\subsection{Spectral analysis}
Our morphological analysis indicates some incongruity between the morphology of the radio lobe and $\gamma$-ray lobe in the north.
Hence, to model the $\gamma$-ray lobe as self-consistently as possible, we used the template generated with the residual map (T1).
Integrating the whole $\gamma$-ray emission observed, we then derived the total flux and index in the $100$ MeV to $30$ GeV energy range.
For the north lobe the integral HE flux is $(0.93\pm 0.09) \times 10^{-7} \rm ph~cm^{-2}s^{-1}$ and the photon index is $2.24\pm 0.08$, while
for the south lobe we find $(1.4\pm  0.2) \times 10^{-7} \rm ph~cm^{-2}s^{-1}$ and $2.57\pm 0.07$, respectively.  The core region of Cen A
has a flux $(1.4\pm  0.2) \times 10^{-7} \rm ph~cm^{-2}s^{-1}$ while the photon index is $2.7 \pm 0.1$. The results are summarized
in Table~2, where the subscripts 3a and 10m refer to the three-year data (analyzed here) and the ten-month data ( reported in Abdo et al. 2010a),
respectively. We find that the flux and photon indices in the T2 templates are similar to the ten-month data. On the other hand, the analysis
using the T1 template results in a harder spectrum for the north lobe. \\
To derive the spectral energy distribution (SED) we divided the energy range into logarithmically spaced bands and applied {\it gtlike} in each
of these bands. Only the energy bins for which a signal was detected with a significance of at least $2 \sigma$ wereconsidered, while an upper
limit was calculated for those below. As a result, there are seven bins in the SED for the south lobe. The SED is shown in Fig.~4.\\
To clarify the origin of the $\gamma$-ray emission, we evaluated the spectrum in different parts of each lobe. To this end, we divided each
lobe into two parts (see Fig.~5) and used {\it gtlike} to evaluate the spectrum. In the south lobe the resulting photon index is $2.8\pm 0.2$
near the Cen A core and $2.3\pm 0.1$ far away from the core.  Unfortunately, the statistics are still not high enough to claim a clear
hardening of the spectrum. For the northern lobe, both parts appear to be consistent with values of $2.2\pm 0.2$. The spectra for the four
regions are shown in Fig.~6 and Fig.~7.
\begin{table}[htbp]
\caption{Fluxes and spectra of the lobes} \label{tab:2} \centering
\begin{tabular}{lllllll}
\hline
Source Name &\vline $\Phi_{3a}(\rm T1)$&\vline$\Gamma_{3a}(\rm T1)$&\vline$\Phi_{10m}$&\vline$\Gamma_{10m}$&\vline$\Phi_{3a}(\rm T2)$&\vline$\Gamma_{3a}(\rm T2)$\\
\hline
South Lobe &\vline$1.43\pm 0.15$&\vline$2.57\pm 0.07$&\vline$1.09\pm 0.24$&\vline$2.60\pm 0.15$&\vline$1.40\pm 0.15$&\vline$2.56\pm 0.08$\\
\hline
North Lobe &\vline $0.93\pm 0.09$&\vline$2.24\pm 0.08$&\vline$0.77\pm 0.20$&\vline$2.52\pm 0.16$&\vline$0.64\pm 0.15$&\vline$2.56\pm 0.08$\\
\hline
\end{tabular}
\caption*{$\Phi$ is the integral flux (100 MeV to 30 GeV) in units of $10^{-7} \rm ph \cdot cm^{-2}s^{-1}$ and $\Gamma$ is the photon index. The
subscripts "3a" and "10m" refer to the three-year data analyzed here and to the ten-month results  (based on a WMAP template) reported in Abdo et al.
(2010a), respectively.}
\end{table}
\begin{figure}[htbp]
\centering
\subfloat[LAT counts map of the $14^{\circ}\times14^{\circ}$ ROI.]{\includegraphics[width=0.45\textwidth]{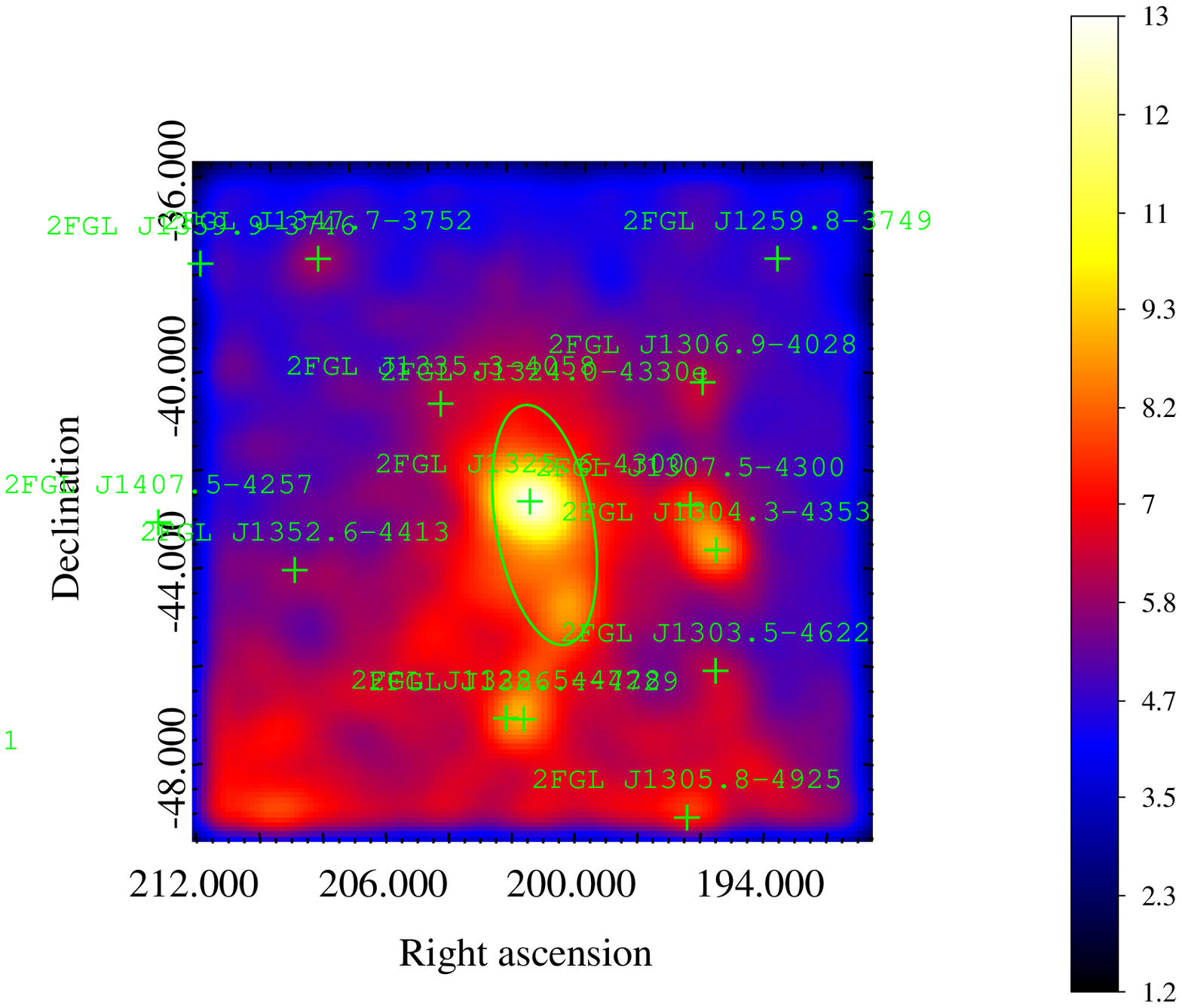}}
\subfloat[Excess map after background subtraction.]{\includegraphics[width=0.45\textwidth]{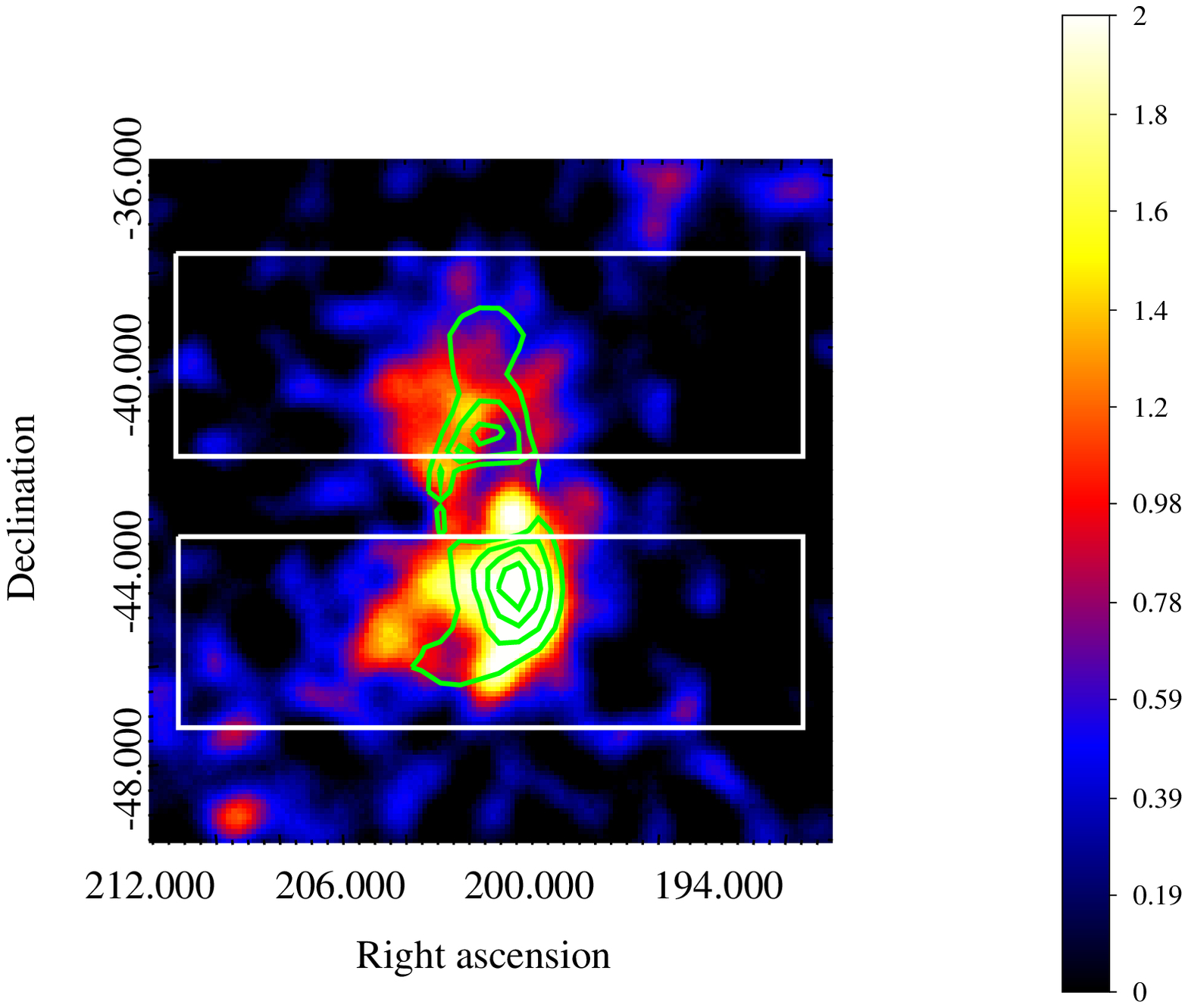}}\\
\subfloat[Residual map for template T1.]{\includegraphics[width=0.45\textwidth]{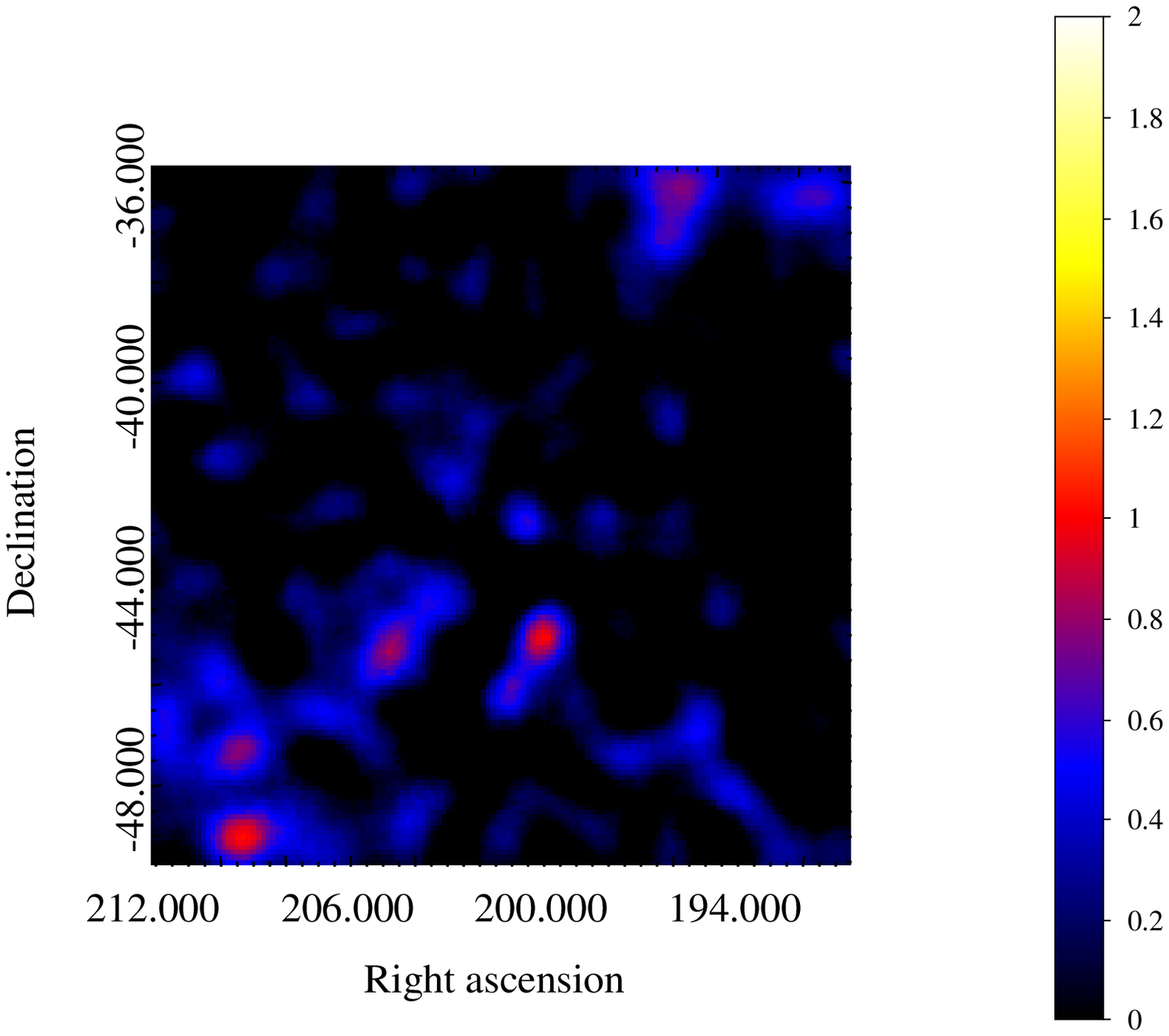}}
\subfloat[Residual map for template T2.]{\includegraphics[width=0.45\textwidth]{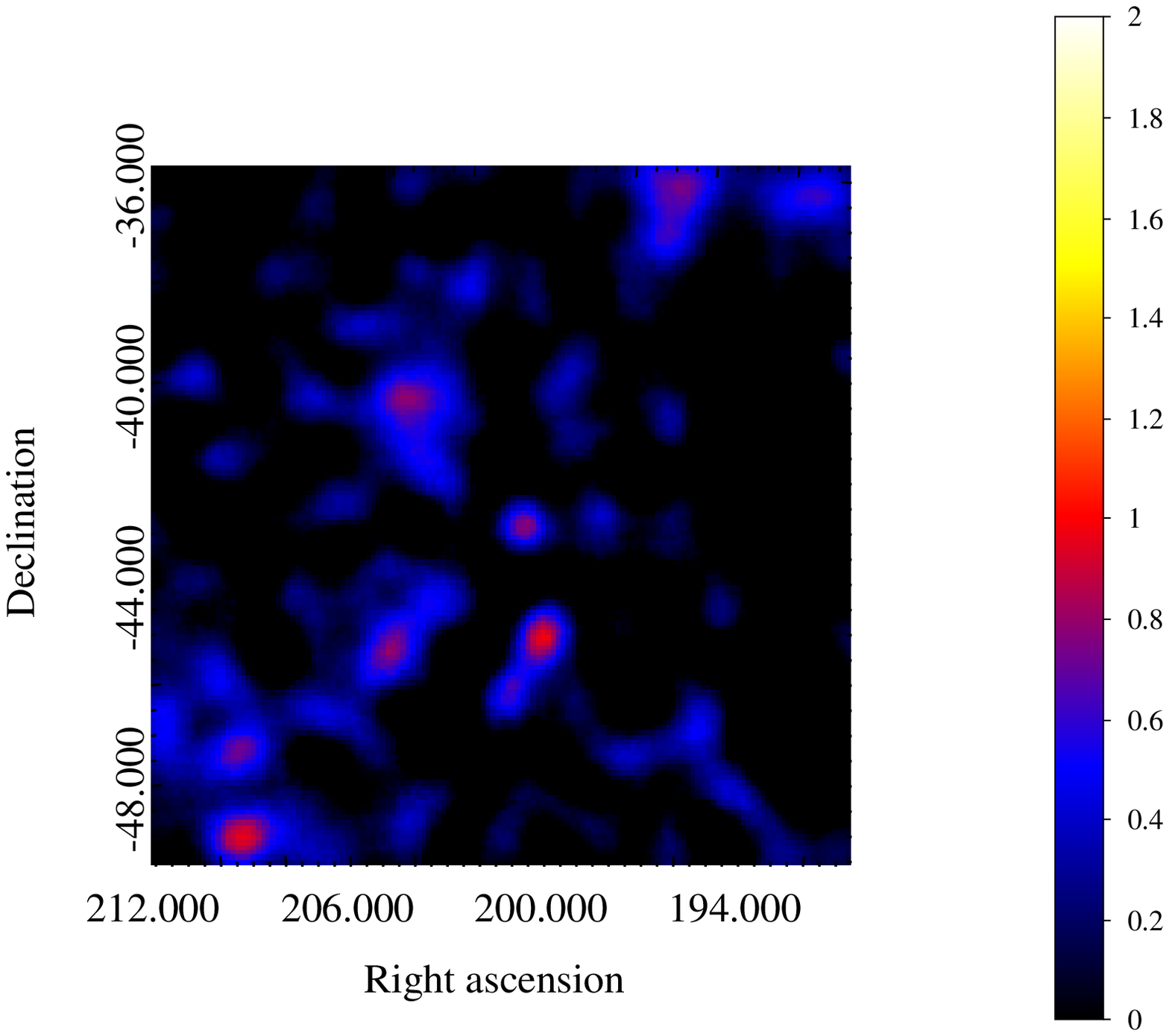}}

\caption{The different maps for the Cen A region: (a) LAT counts map of the $14^{\circ}\times14^{\circ}$ region of interest (ROI) around the position
of Cen A. The counts map is smoothed with a Gaussian of kernel $0.8^{\circ}$. The green crosses mark the position of the 2FGL point-like sources.
(b) Excess map after subtraction of diffuse background, point-like sources and Cen A core. The contours are WMAP radio lobe contours, while the
white boxes represent the projection regions discussed in Sec.~2.1. (c) Residual map using template T1 for the lobes. (d) Residual map using the
radio template T2 for the lobes.}
\end{figure}

\section{WMAP data}
In Sec.~2 we found evidence that the HE $\gamma$-ray emission regions do not coincide well with the radio lobes, especially for the north lobe.
To correctly compare the $\gamma$-ray emission with the radio emission, an analysis of the radio data for the same region is required, rather than
simply for the north radio lobe itself.
We performed this analysis following the method described in Hardcastle et al. (2009). Seven years of WMAP data were analyzed (\cite{komatsu11}).
The WMAP maps in all five bands are convolved with a $0.83^{\circ}$ Gaussian to obtain a  similar resolution for all bands. The internal linear combination
(ILC) cosmic microwave background (CMB) map was treated as background and subtracted from all maps. The intensity maps (in $\rm mK$) were
converted to flux maps (in $\rm Jy/beam$) (\cite{page03}) and integrated in the region defined by the north $\gamma$-ray lobe to obtain the total
flux. Flux values almost twice as high were obtained when the $\gamma$-ray region was used. To cross-check our analysis method, we also derived
the flux of the north radio lobe for the same region as used in Hardcastle et al. (2009). All results are summarized in Tab. 4. Our results for the north
radio lobe are compatible within errors with those obtained by Hardcastle et al. (2009) (shown in parenthesis in Tab. 3).
\begin{table}[htbp]
\caption{Radio flux for the north lobe (in $10^{11}\rm Jy\cdot Hz$) from WMAP data.} \label{tab:4} \centering
\begin{tabular}{lll}
\hline
band  &\vline north $\gamma$-ray Lobe &\vline north radio Lobe\\
\hline
K(22.5GHz)&\vline $5.34\pm0.37$ &\vline $2.71\pm0.19$ ($2.61\pm 0.25$) \\
\hline
Ka(33GHz) &\vline  $5.24\pm 0.41$&\vline $2.55\pm 0.22$ ($2.50\pm 0.24$)\\
\hline
Q(41GHz) &\vline  $5.53\pm 0.56$&\vline $2.94\pm0.29$ ($2.58\pm 0.25$)\\
\hline
V(61GHz) &\vline  $7.65\pm 1.19$&\vline$<4.13$\\
\hline
W(94GHz) &\vline  $<25.6$&\vline$<9.43$\\
\hline
\end{tabular}
\end{table}
Note that in the V and the W band the signal is detected at less than the 5 $\sigma$ level, so we only give upper limits here.

\section{The origin of the non-thermal lobe emission}
Using the WMAP and Fermi-LAT results reported here, we can characterize the spectral energy distributions  for the north
and the south lobe. While the radio emission is usually taken to be caused by electron synchrotron emission, the
origin of the HE $\gamma$-ray emission could in principle be related to both leptonic (inverse-Compton scattering) and hadronic
(e.g., pp-interaction) processes. In the following we discuss possible constraints for the underlying radiation mechanism as
imposed by the observed SEDs.

\subsection{Inverse-Compton origin of $\gamma$-rays}
Both the HE $\gamma$-ray and the radio emission could be accounted for in a leptonic scenario. In the simplest version, a
single population of electrons $N(\gamma,t)$ is used to model the SED through synchrotron and inverse-Compton emission,
with particle acceleration being implicitly treated by an effective injection term $Q=Q(\gamma,t)$. The latter allows us to
distinguish acceleration caused by, e.g., multiple shocks or stochastic processes (e.g., \cite{osullivan09}) from emission, and enables
a straightforward interpretation. The kinetic equation describing the energetic and temporal evolution of the radiating electrons
can then be written as
\begin{equation}
\frac{\partial N}{\partial t}=\frac{\partial}{\partial \gamma}\left(P\:N \right)-\frac{N}{\tau_{esc}}+Q\:,
\end{equation}
where $P=P(\gamma)=-\frac{d\gamma}{dt}$ is the (time-independent) energy loss rate and $\tau_{\rm esc}$ is the characteristic
escape time. For negligible escape (as appropriate here, given the large size of the $\gamma$-ray emitting region) and  quasi
continuous injection (considered as a suitable first-order approximation given the short lifetime of TeV electrons and the scales
of the lobes) $Q(\gamma,t)=Q(\gamma)$, the solution of the kinetic equation becomes
\begin{equation}\label{Nsolution}
N(\gamma,t)=\frac{1}{P(\gamma)} \int^{\gamma_{0}}_{\gamma}Q(\gamma)d\gamma\;,
\end{equation}
where $\gamma_{0}$ is found by solving the characteristic equation for a given epoch $t$, $t=\int^{\gamma_{0}}_{\gamma}
\frac{d\gamma}{P(\gamma)}$ (e.g., \cite{atoyan99}). If synchrotron or inverse-Compton (Thomson) losses ($P(\gamma)=a\:
\gamma^{2}$) provide the dominant loss channel, then $\gamma_0=\gamma/(1-a\gamma t)$, so that at the energy $\gamma_{\rm br}
=\frac{1}{a\: t}$ the stationary power-law electron injection spectrum steepens by a factor of $1$ (cooling break) due to radiative
losses, naturally generating a broken power-law.\\
We used the above particle distribution described in eq. (\ref{Nsolution}) for a representation of the observed lobe SEDs. The
magnetic field strength B, the maximum electron energy $\gamma_{\rm max}$ and the epoch time t were left as free parameters
to model the data. Klein-Nishina (KN) effects on the inverse-Compton-scattered HE spectrum were taken into account (following
\cite{aharonian81}).\\
Fig. 8 shows the SED results obtained for the north and south lobes. The HE part of both spectra can be described by a
power-law with photon index $\Gamma_{\gamma} \simeq 2.2$ and $2.6$ for the north and the south lobe, respectively.
At low energies, the south lobe spectrum shows a synchrotron peak at about $5$ GHz, while the north lobe is well described
by a power-law with an index $>2$. Note that if one would use a simple power-law electron injection spectrum $Q(\gamma)
\propto \gamma^{-\alpha}$, evolving in time with a cooling break, to describe the HE $\gamma$-ray spectrum, a power index
$\alpha=3.2$ would be required for the south lobe. Yet, assuming that the same electron population is responsible for both
the radio-synchrotron and HE inverse-Compton emission, such a value would be in conflict with the results obtained from
the WMAP data analysis, indicating an electron population with power-law index $\alpha\simeq 2$ based on the detected
synchrotron emission. As it turns out, however, this problem could be accommodated by considering a more natural spectral
input shape, e.g., an electron injection spectrum with an exponential cut-off
\begin{equation}
Q(\gamma)=Q_{0}\:\gamma^{-\alpha}\exp\left(-\frac{\gamma}{\gamma_{\rm max}}\right)\:,
\end{equation}
where the constant $Q_{0}$ can be obtained from the normalization to the injection power $L=m_ec^{2} \int Q(\gamma)\;
\gamma\;d\gamma$.\\
The age of the giant lobe emission, and associated with this, the duration of particle acceleration activity, is somewhat
uncertain. Dynamical arguments suggest a lower limit $>10^7$ yr for the giant radio lobes, while synchrotron spectral
aging arguments indicate an age $\simlt 3 \times 10^7$ yr (e.g., \cite{israel98, alvarez00, hardcastle09}). The observed
GeV extension in itself would already imply an extreme lower limit of $R/c >10^6$ yr. In the following we therefore discuss
the SED implications for an epoch time $t$ between $10^{7}$ yr and $10^{8}$ yr. As it turns out, the modeling of the GeV
data provides support for a maximum lobe age of $\sim 8 \times 10^7$ yr.\\
For the south lobe, the radio data suggest a break frequency $\nu_{\rm br}=5$~GHz above which the spectrum drops abruptly.
The break in the synchrotron spectrum is related to the break in the electron spectrum via $\nu_{ \rm br}=1.3\:\gamma_{\rm br}^{2}\:
B_{1\mu G}$~Hz. In principle, a change in the spectral shape of the electron population might be caused by cooling effects
or/and the existence of a maximum energy for the electron population. For a minimum epoch time $t_{\rm min}=10^{7}$ yr,
cooling would affect the synchrotron spectrum at frequency $\approx 80 B_{1\mu G}$~GHz, much higher than inferred
from the radio data. Therefore, to obtain a break at $5$ GHz in the south lobe, a maximum energy in the electron population
($\gamma_{\rm max}$), lower than $\gamma_{\rm br}$ defined by $t=t_{\rm min}$ would be needed. On the other hand,
for a maximum epoch time $t_{\rm max}=8\times 10^{7}$ yr, the power-law spectral index changes at frequency $\simeq
1 B_{1\mu G}$ GHz, providing a satisfactory agreement with the radio data. In this case the maximum electron energy is
obtained from the radio data above the break frequency 5~GHz. Results for the considered minimum and maximum epoch
time, and for a fixed power-law electron index $\alpha=2$ are illustrated in Figs. 8 and 9. Note that for $B\leq 3 \mu$G,
the energy loss rate $P(\gamma)$ is dominated by the IC channel only, so that the results of the calculations are quite
robust.\\
Figure 8 shows a representation of the SED for the north and the south lobe, respectively, using the parameters $t_{\rm min}
=10^{7}$ yr and $\gamma_{\rm max}=1.5\times 10^5$. The dashed line shows the HE contribution produced by inverse-Compton
scattering of cosmic microwave background  photons by relativistic electrons within the lobes. In this case
the resulting $\gamma$-ray flux can only describe the first two data points and then drops rapidly.
Consequently, to be able to account for the observed HE spectrum, extragalactic background light  photons need to
be included in addition to CMB photons (see dot-dashed line in Figure 8 ).  Upscattering of infrared-to-optical EBL photons
was already required in the stationary leptonic model reported in the original Fermi paper (\cite{abdo10a}). In our approach
we adopt the model by \cite{franceschini08} to evaluate this EBL contribution.
The solid line in Fig. 8 represents the total (CMB+EBL) inverse-Compton contribution. The maximum total energy of electrons
in both lobes is found to be $\sim 2\times10^{58}$ erg  and the energy in the magnetic fields is roughly $25\%$ of this.
Dividing the total energy by the epoch time $10^7$ yr would imply a mean kinetic power of the jets inflating the lobes of $\simeq
7\times 10^{43}$ erg/s, roughly two orders of magnitude lower than the Eddington power inferred for the black hole mass in
Cen A, yet somewhat above the estimated power of the kpc-scale jet in the current epoch of jet activity (\cite{croston09}).
 This could indicate that the jet was more powerful in the past. Obviously, the requirement on the mean jet power can be
significantly reduced if one assumes an older age of the lobes.\\
Figure 9 shows a representation of the SED for an epoch time $t_{\rm max}=8 \times 10^{7}$ yr, with a maximum electron
Lorentz factor $\gamma_{\rm max}= 2.5 \times 10^{6}$ and $1.5 \times 10^{6}$ for the north lobe and the south
lobe, respectively. Note that in this case the contribution by inverse-Compton scattering of CMB photons  alone is sufficient
to account for the observed HE spectrum (see the solid line in Figure 9). The inverse-Compton contribution of EBL photons
only becomes important at higher energies (see the dot-dashed line in Fig. 9). On the other hand, for an epoch time $t$
exceeding $t_{\rm max}=8\times 10^{7}$ yr, the high-energy part of the SED would no longer be consistent with the data (see
the dashed line in Fig. 9 for $t=10^{8}$ yr). This could be interpreted as additional evidence for a finite age $< 10^8$ yr of the
lobes. The maximum total energy of electrons in both lobes is found to be $\approx6\times10^{57}$ erg,  with the total
energy in particles and fields comparable to the $10^7$yr-case, thus requiring only a relatively modest mean kinetic jet power
of $\sim 10^{43}$ erg/s.

\subsection{Hadronic $\gamma$-rays?}
Once protons are efficiently injected, they are likely to remain energetic since the cooling time for pp-interactions is
$t_{\rm pp} \approx 10^{15}(n/1\;cm^{-3})^{-1}$ s. High-energy protons interacting with the ambient low-density plasma
can then produce daughter mesons and the $\pi^{0}$ component decays into two $\gamma$-rays. The data reported
here allow us to derive an upper limit on the energetic protons contained in the lobes of Cen A. As before, we use a
power-law proton distribution with an exponential cut-off, i.e.,
$$N(\gamma_p)=N_{0}\;\gamma_p^{-\alpha } \exp\left(-\frac{\gamma_p}{\gamma_{\rm max}}\right)$$
where the constant $N_{0}$ can be expressed in terms of the total proton energy $W_{p}=m_p c^2 \int \gamma_p\;
N(\gamma_p)\:d\gamma_p$. Current estimates for the thermal plasma density in the giant radio lobes of Cen A suggest
a value in the range $n \simeq (10^{-5}-10^{-4})$ cm$^{-3}$ (e.g., \cite{isobe01, feain09}). We used $n=10^{-4}$ cm$^{-3}$
for the model representation shown in dotted line in Fig. 9. In both lobes, the power-law index of the proton population is
$\alpha=2.1$, and the high-energy cut-off is $E_{\rm max} \simeq 55$ GeV. The maximum total energy $W_{p}$ is
proportional to the gas number density $n$, so that $W_{p} \simeq 10^{61}(n/10^{-4}\;{\rm cm}^{-3})^{-1}$ erg, obtained
here, should be considered as an upper limit. In principle, protons could be accumulated over the whole evolutionary
timescale of the lobes. For a long timescale of $\geq 10^9$ yr, an average injection power $\leq 3 \times 10^{44}$ erg/s and
a mean cosmic-ray diffusion coefficient of $D \sim R^2/t \simlt 3 \times 10^{30} (R/100~\rm{kpc})^2$ cm$^2$/s would
be needed.

\section{Discussion and conclusion}
Results based on an detailed analysis of 3 yr of Fermi-LAT data on the giant radio lobes of Cen A are described in this paper.
We have shown that the detection of the HE lobes with a significance more than twice as high as reported before (i.e., with
more than $10$ and $20 \sigma$ for the northern and the southern lobe, respectively) allows a better determination of
their spectral features and morphology. A comparison of the Fermi-LAT data with WMAP data indicates that the HE $\gamma$-ray
emission regions do not fully coincide with the radio lobes. There is of course no a priori reason for them to coincide. The
results reported here particularly support  a substantial HE $\gamma$-ray extension beyond the WMAP radio
image for the northern lobe of Cen A. We have reconstructed the SED based on data from the same emission region.
A satisfactory representation is possible in a time-dependent leptonic scenario with radiative cooling taken into account
self-consistently and injection described by a single power-law with exponential cut-off. The results imply a finite age $<10^8$
yr of the lobes and a mean magnetic field strength $B\simlt 1 \mu$G. While for lobe lifetimes on the order of $8\times 10^7$ yr,
inverse-Compton up-scattering of CMB photons alone would be sufficient to account for the observed HE spectrum, up-scattering
of EBL photons is needed for shorter lobe lifetimes. In a leptonic framework, the HE emission directly traces (via
inverse-Compton scattering) the underlying relativistic electron distribution and thereby provides a spatial diagnostic tool.
The radio emission, arising from synchrotron radiation, on the other hand also traces the magnetic field distribution. Together,
the HE $\gamma$-ray and the radio emission thus offer important insights into the physical conditions of the source. That the HE emission seems extended beyond the radio image could then be interpreted as caused by a change in the magnetic
field characterizing the region. This would imply that our quasi-homogeneous SED model for the  HE lobes can only serve
as a first-order approximation and that more detailed scenarios need to be constructed to fully describe the data. This also applies
to the need of incorporating electon re-acceleration self-consistently. Extended HE emission could in principle also be related
to a contribution from hadronic processes. The cooling timescales for protons appear much more favorable. On the other hand,
both the spectral shape of the lobes and the required energetics seem to disfavor pp-interaction processes as sole contributor.
One of the insights emerging from the present paper is the need for a more detailed theoretical SED approach to be able
to take full advantage of the current observational capabilities.

\acknowledgement
 Constructive comments by the anonymous referee are gratefully acknowledged.

\begin{figure}
\centering
\includegraphics[width=120mm,angle=0]{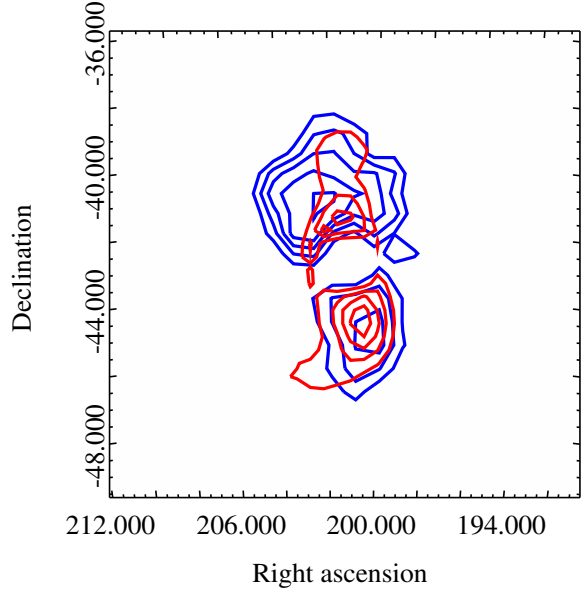}
\centering
\caption{Two templates used in the analysis. The blue contours correspond to T1 and the red to T2.}
\end{figure}
\begin{figure}[!ht]
\centering
\subfloat[Projection of the south lobe]{\includegraphics[width=0.45\textwidth]{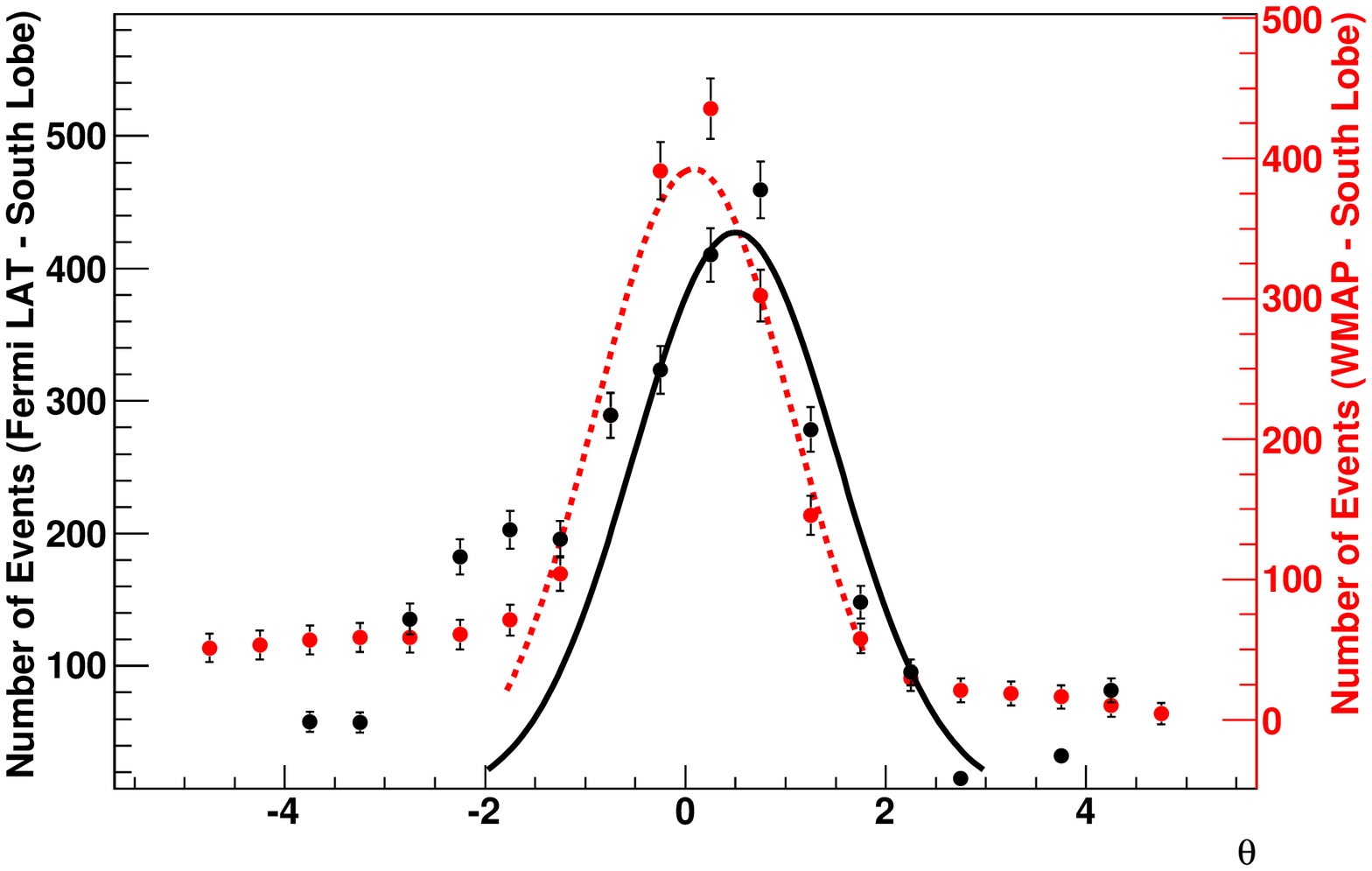}}
\subfloat[Projection of the north lobe]{\includegraphics[width=0.45\textwidth]{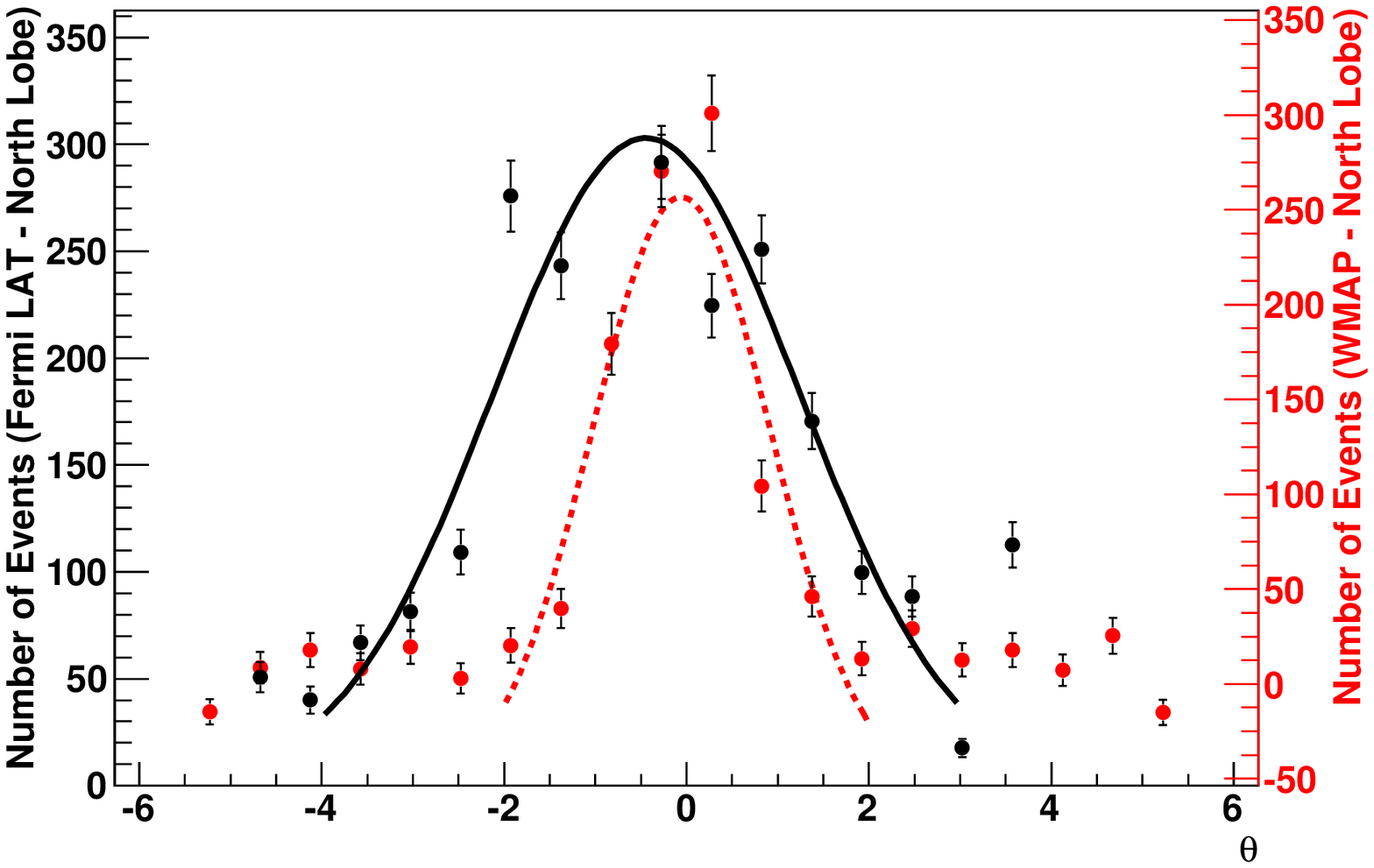}}
\centering
\caption{Projection of the rectangular region shown in Fig. 1(b) for both lobes. The curves are Gaussian fits, with WMAP
data in red and Fermi-LAT data in black.  Units for the positional (x) axis are degrees.}
\end{figure}
\begin{figure}
\centering
  \includegraphics[width=120mm,angle=0]{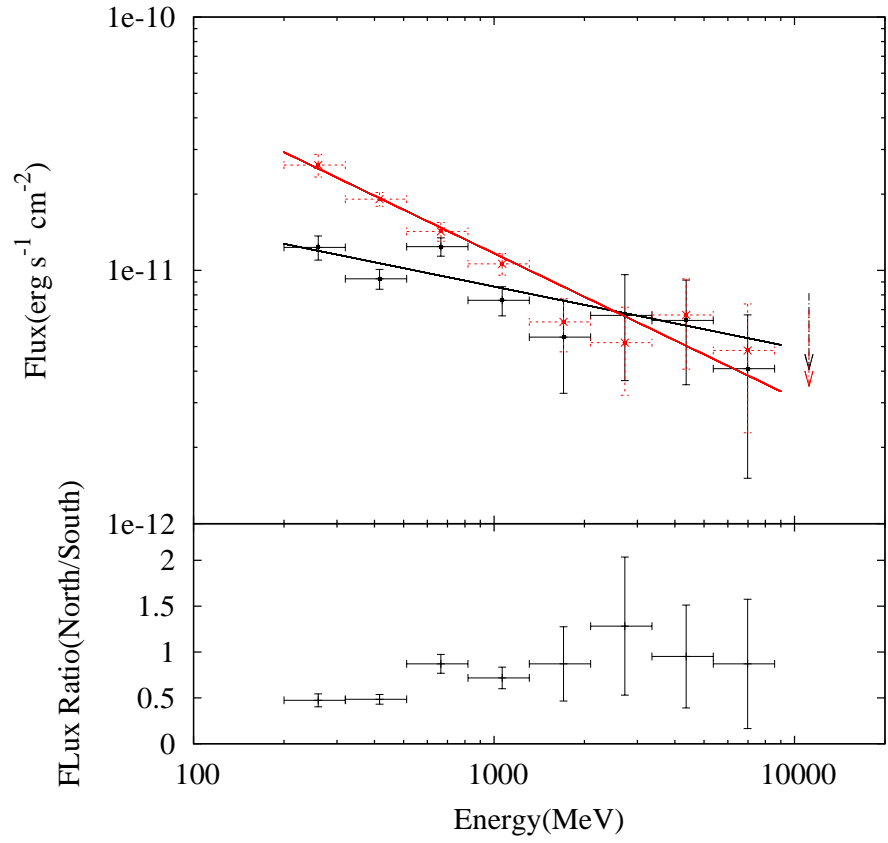}
\caption{SED for template T1. Squares and crosses are for the north lobe and the south lobe, respectively. The ratio of the
fluxes (north/south) are shown in the bottom panel.}
\end{figure}
\begin{figure}
\centering
\includegraphics[width=120mm,angle=0]{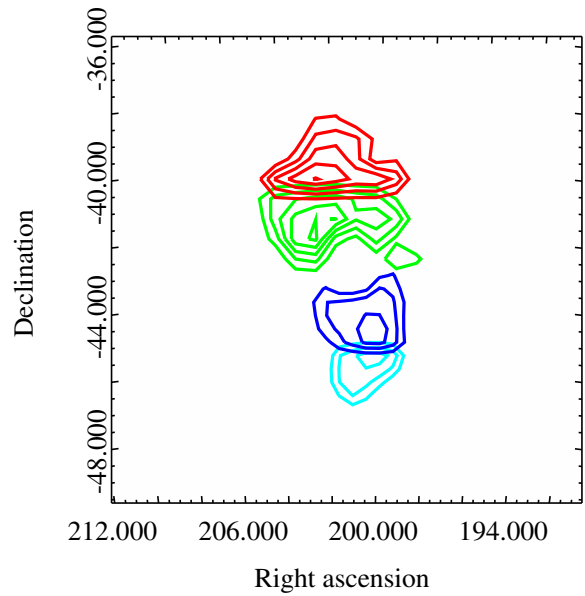}
\caption{Two lobes divided into two parts (near core of Cen~A core and farther away) for spectral comparison.
The contours in different colors represent the different parts.}
\end{figure}
\begin{figure} [h!!]
{\includegraphics[width=85mm,angle=0]{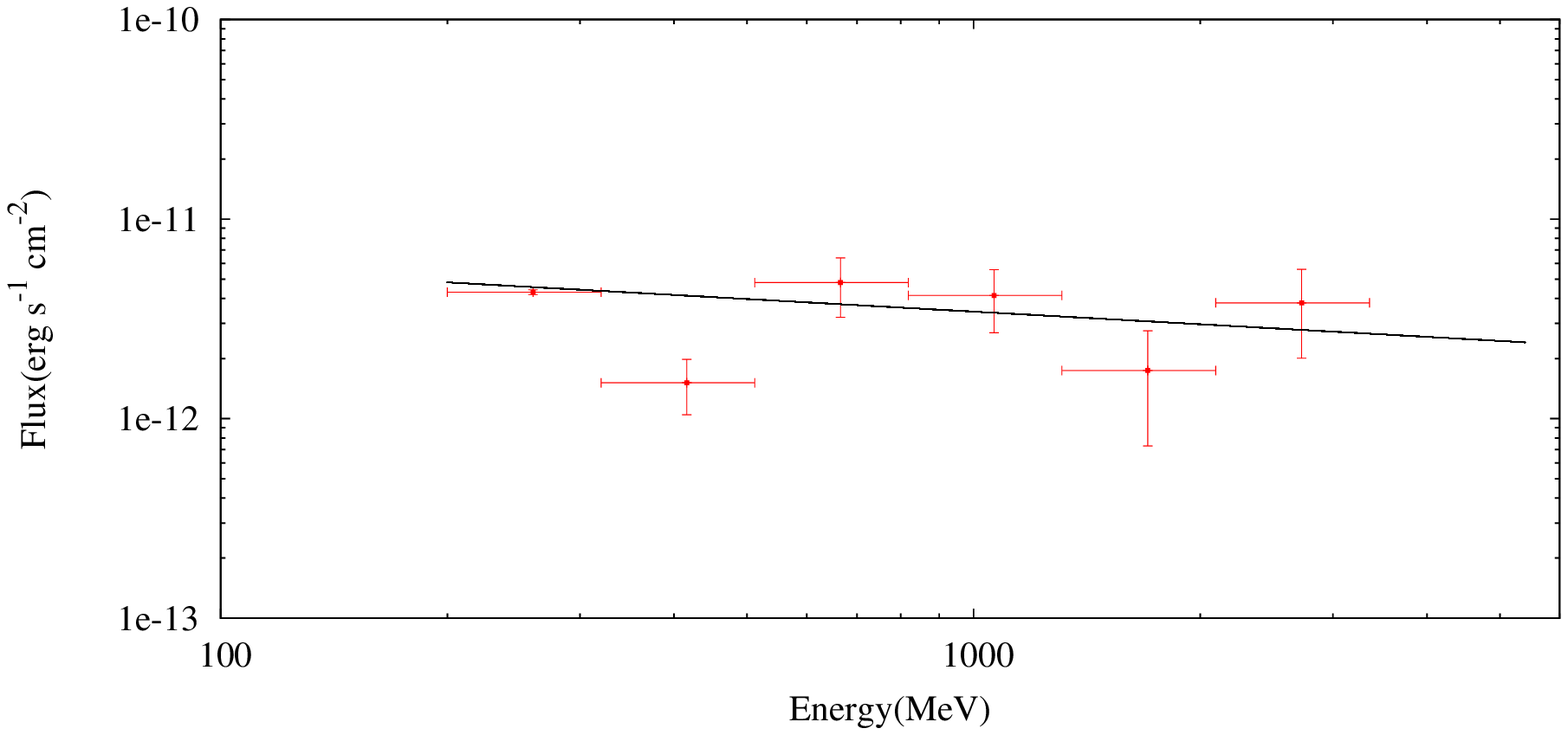}}
{\includegraphics[width=85mm,angle=0]{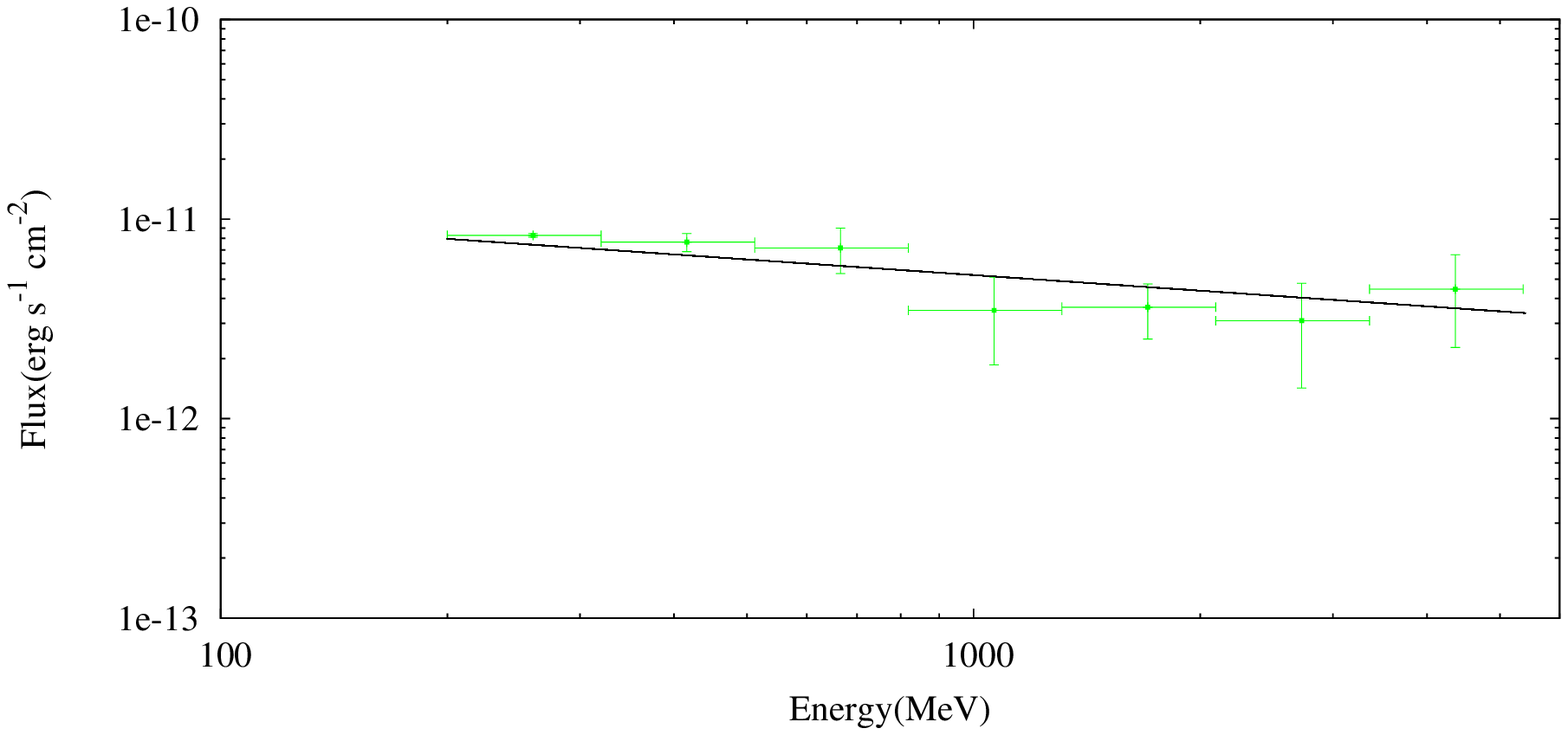}}\\
\caption{SEDs for the corresponding subregions in the north. The left panel is the SED of the subregion far from the core,
while the right panel refers to the subregion near the core. The colors are the same as used in the definition in Fig. 5.}
\end{figure}
\begin{figure} [h!!]
{\includegraphics[width=85mm,angle=0]{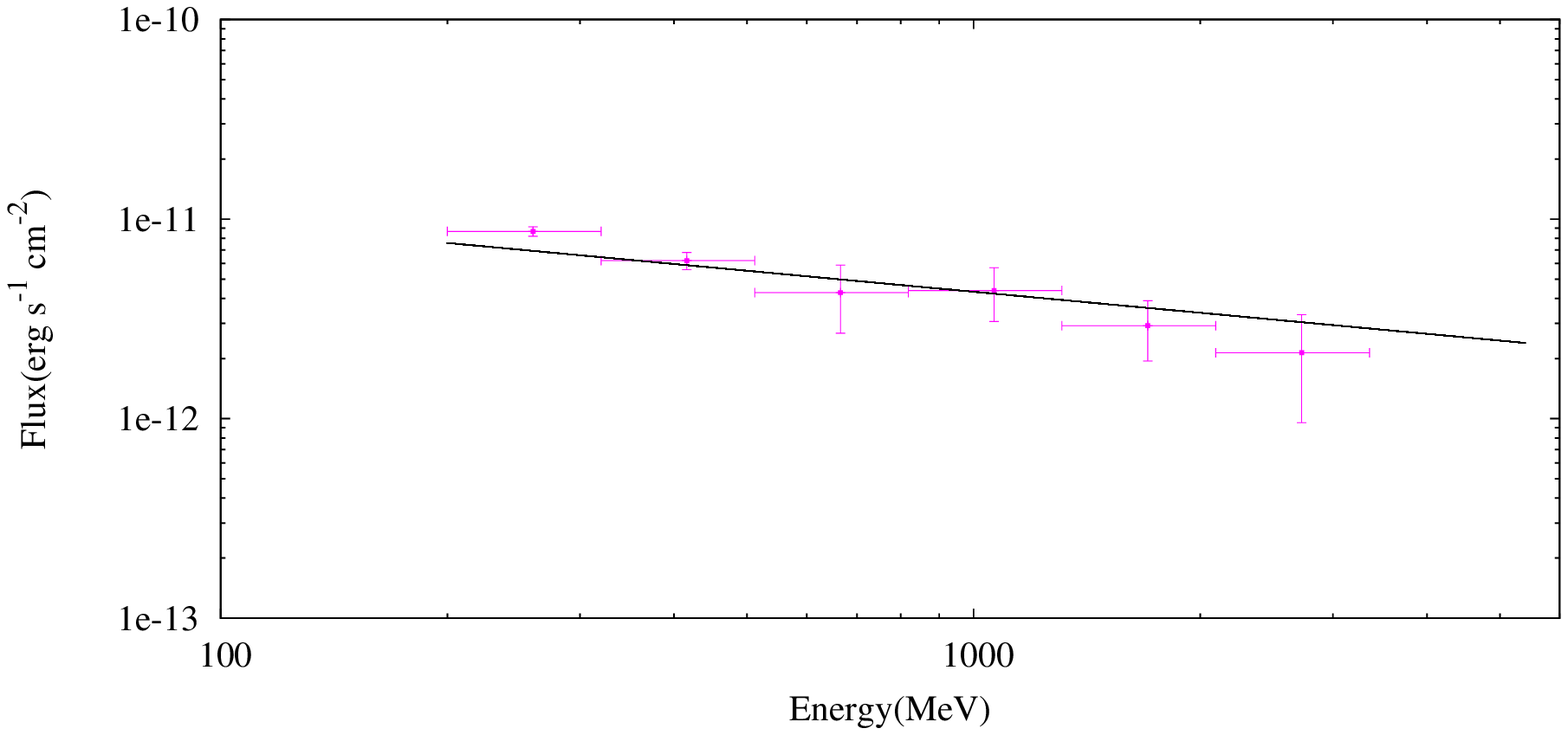}}
{\includegraphics[width=85mm,angle=0]{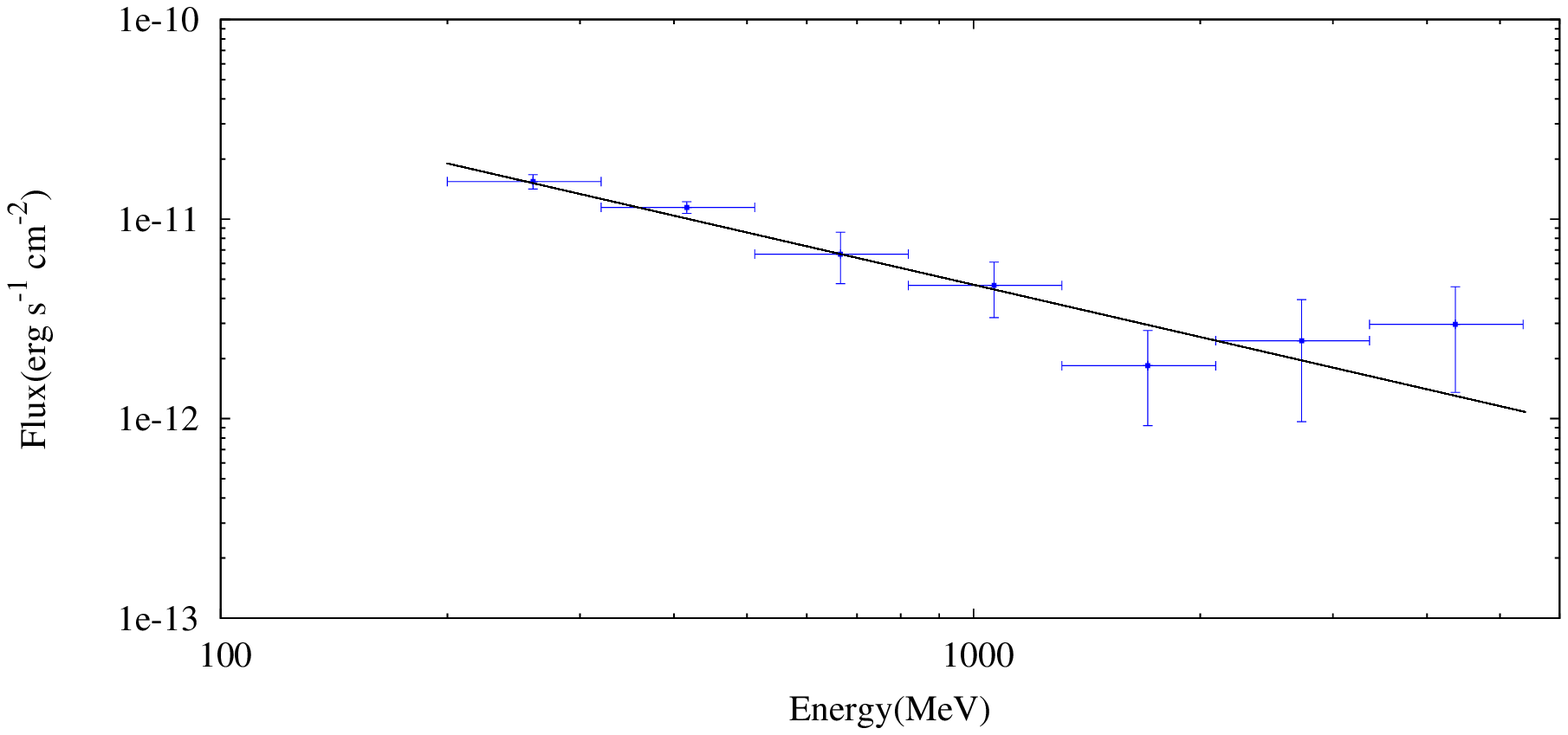}}
\caption{Same as Fig. 6 but for the southern subregions. The left panel is the SED of the subregion far from the core,
while the right panel refers to the subregion near the core.}
\end{figure}
\begin{figure}[!ht]
\centering
\subfloat[South lobe]{\includegraphics[width=0.45\textwidth]{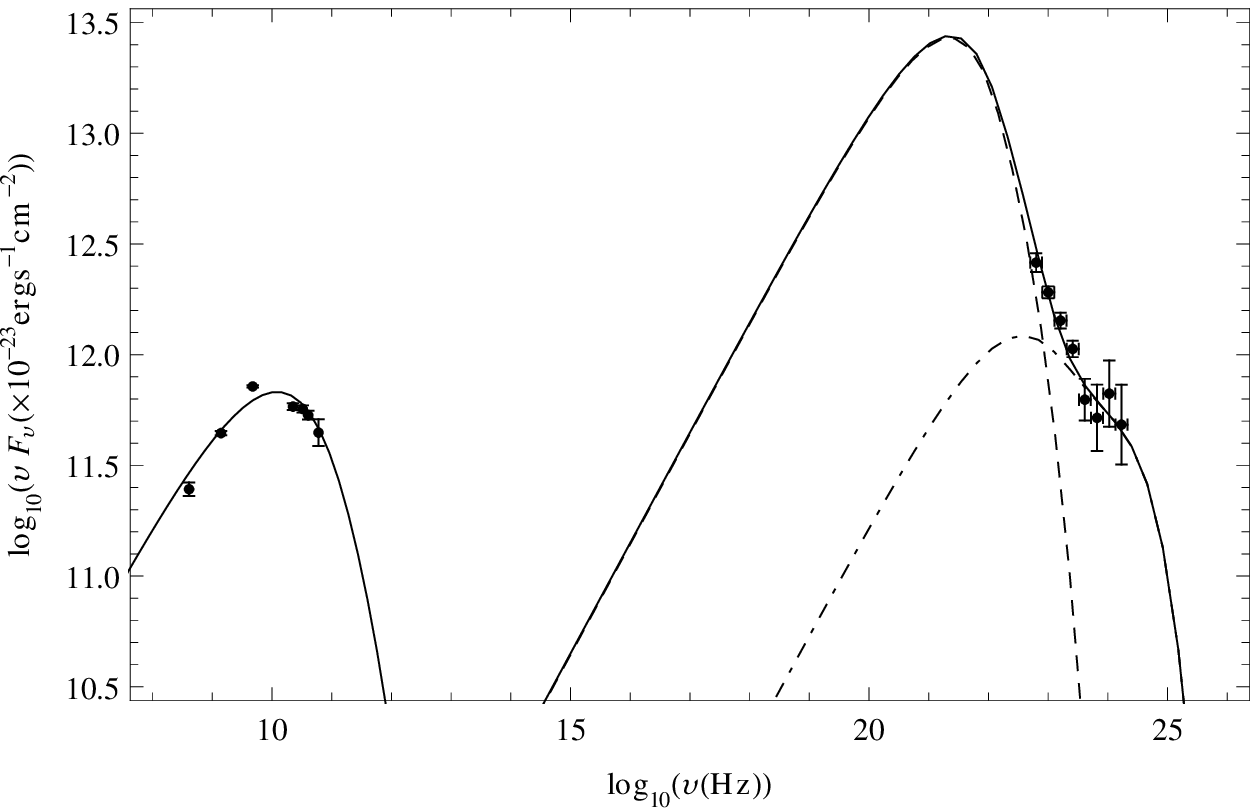}}
\subfloat[$\gamma$-ray excess (lobe) region in the north]{\includegraphics[width=0.45\textwidth]{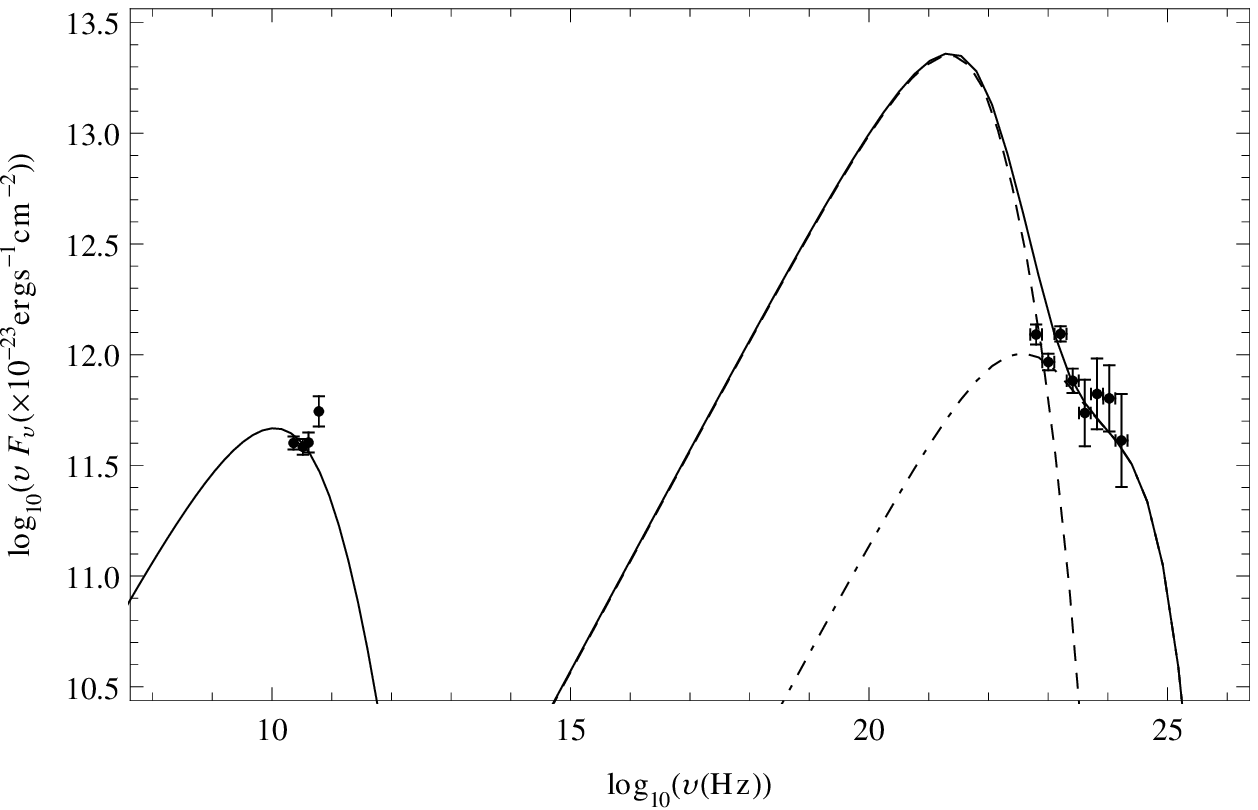}}
\centering
\caption{Synchrotron and inverse-Compton fluxes for $t=10^{7}$ yr. The radio data for the south lobe are from Hardcastle et al.
(2009) (sum of region 4 and region 5 in their Table 1), while the radio data for the north region are from the WMAP analysis
in this paper. The mean magnetic field value $B$ used for the north and the south lobe is $0.39\;\mu G$ and $0.43\;\mu G$,
respectively. The dot-dashed line refers to the IC contribution due to EBL upscattering.}
\end{figure}
\begin{figure}[!ht]
\centering
\subfloat[South lobe]{\includegraphics[width=0.45\textwidth]{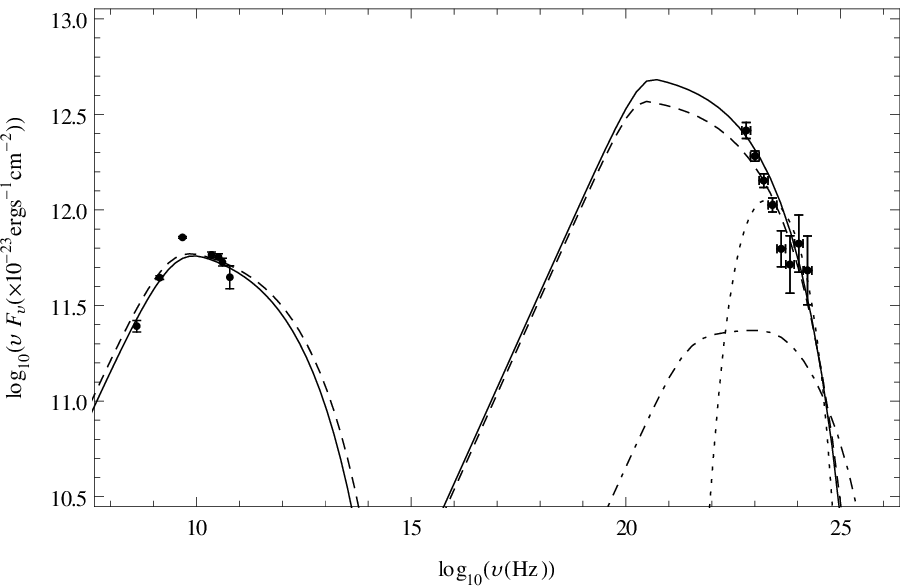}}
\subfloat[$\gamma$-ray excess (lobe) region in the north]{\includegraphics[width=0.45\textwidth]{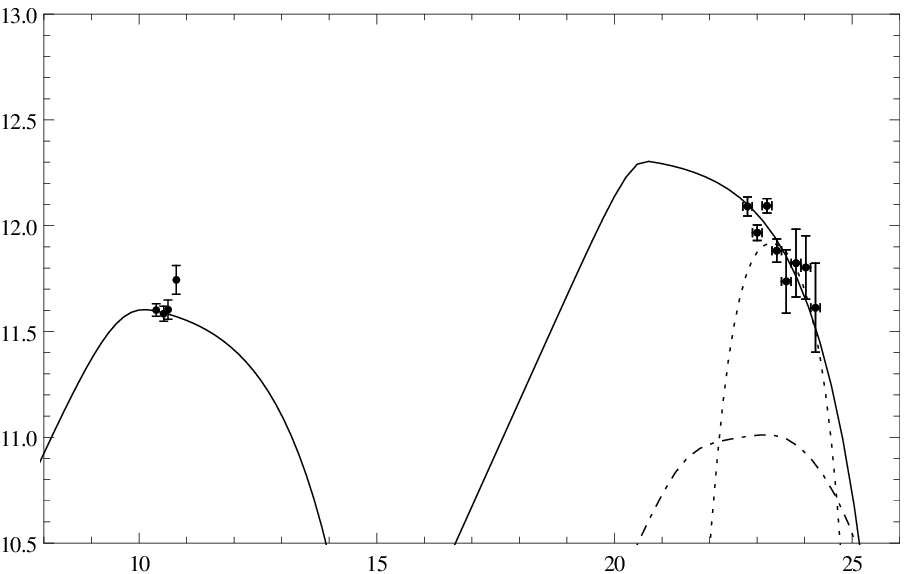}}
\centering
\caption{Synchrotron and inverse-Compton fluxes for $t=8\times10^{7}$ yr. The mean magnetic field value $B$ for the south
lobe and the $\gamma$-ray excess region in the north lobe is $0.91\;\mu G$ and $1.17\;\mu G$,  respectively. The dot-dashed line refers
to the IC contribution due to EBL upscattering. The dashed line (a) shows the result for $t=10^8$ yr. The possible $\gamma$-ray
flux expected from pp-interactions for a thermal gas density $n=10^{-4}\; cm^{-3}$ are also shown (dotted line).}
\end{figure}

\end{document}